**Cryogenic sapphire oscillator with exceptionally high long-term frequency stability**

J.G. Hartnett, C.R. Locke, E.N. Ivanov, M.E. Tobar, P.L. Stanwix

School of Physics, The University of Western Australia

We report on the development of a sapphire cryogenic microwave resonator oscillator with long-term fractional frequency stability of $2 \times 10^{-17}\sqrt{\tau}$ for integration times $\tau > 10^3$ s and a negative drift of about $2.2 \times 10^{-15}$/day. The short-term frequency instability of the oscillator is highly reproducible and also state-of-the-art: $5.6 \times 10^{-16}$ for an integration time of $\tau \approx 20$ s.

Cryogenic sapphire oscillators (CSOs)[1-4] operating at microwave frequencies are capable of exhibiting a fractional frequency instability less than a part in $10^{15}$. Such oscillators are essential for operation of laser cooled atomic clocks at the quantum noise limit,[5] for precision tests of modern physics including the search for possible violations of Lorentz invariance, by comparing a CSO with a hydrogen-maser[6] or a CSO operating on the beat of same mode in a pair of resonators with orthogonal orientation,[7] and for searching for drift in the fine structure constants, which use fountain clocks with a CSO as the interrogation oscillator.[8,9]

In this letter we outline the design and performance of the upgraded version of our CSO. The CSO has improved short term fractional frequency stability, by a factor of 40 compared to the 1990 version of the CSO,[10] but by a factor of 5 on the 1995 version,[11] yet by a factor of 10 in the long term performance of the latter.

A highly stable CSO has been constructed based on a cylindrical resonator made from the Crystal Systems HEMEX grade single-crystal sapphire. The rotational and crystal axes of the sapphire cylinder were aligned within one degree. The sapphire cylinder was optically polished on all surfaces and featured coaxial spindles at both ends. The bottom longer spindle was used for clamping the sapphire inside the metal cavity to minimize the stress translated to the circumference of the sapphire cylinder where the density of the electromagnetic field was highest.[12,13] The top spindle was used for holding the sapphire during the cleaning only (no pressure was applied to the top spindle from the metal lid in the cavity assembly). The design of the shape of the main body has been previously discussed.[1] It was found that the cylinder design[14] shown in fig. 1, both minimizes the stress translated to the main body and reduces the spurious mode density in the sapphire.

Two types of electromagnetic modes of Whispering Gallery (WG) were excited in the sapphire resonator: WGH (quasi-TM) and WGE (quasi-TE).[15] The highest Q-factor ($1.7 \times 10^9$ at 4.2 K) was measured at the fundamental WGH-mode with the azimuthal number $m = 16$. This mode exhibited a turning point in the frequency-temperature dependence at 7.2319 K due to the combined effects of the sapphire lattice expansion and the Curie Law susceptibility of residual paramagnetic ions.[16] The fractional curvature of the frequency-temperature dependence at the turning point was $2.7 \times 10^{-9}$ K$^{-2}$. The Q-factor of the sapphire loaded cavity resonator was $7 \times 10^8$ at the turnover temperature. Resonator primary and secondary port coupling coefficients at 7.23 K were set to be 1.06 and 0.1, respectively.

The operating mode, WGH$_{16,0,0}$ at 11.200 GHz, is approximately 10 MHz away from the closest spurious mode of the sapphire loaded cavity resonator. The weak coupling of the operating mode to the cavity-like modes (highly tunable with temperature) minimized its frequency-temperature tuning coefficient and was largely responsible for the greatly reduced frequency drift of the CSO (see the details below). The size of the sapphire cylinder was chosen to place the operating mode away from the electron spin resonance frequency due to possible chromium (11.45 GHz) or iron (12.03 GHz) impurities in order to preserve the mode high Q-factor.

The resonator was sealed inside an evacuated stainless steel vacuum can, which in turn was sealed inside another stainless steel vacuum can that was cooled to 4.2 K inside a 250 liter liquid helium dewar, which has sufficient capacity to stay cold without refilling for 30 days. The microwave energy was coupled into the resonator through stainless steel coaxial lines running from a room temperature loop oscillator.

The loop oscillator[10] was based on an Endwave JCA812-5001 microwave amplifier with a small signal gain of 47dB. This amplifier was capable of delivering ~1mW of microwave power to the primary port of the cryogenic resonator. The oscillator frequency was tightly locked to the chosen mode of the cryogenic resonator using the Pound frequency stabilization technique.[17] This involved rapid phase modulation of the microwave signal incident on the cryogenic resonator followed by synchronous demodulation of the reflected microwave signal.

The error voltage from the output of the synchronous demodulator was applied (after appropriate filtering) to a voltage controlled phase shifter in the microwave sustaining loop. This altered the phase delay around the microwave loop steering the oscillator to the resonance frequency of the sapphire loaded cavity.

The temperature of the cryogenic resonator was maintained within a 100 μK offset of the frequency-temperature turnover point (7.2319 K) using a Lakeshore 340 temperature controller and a CGR-1-2000 carbon glass sensor.

Fluctuations of microwave power dissipated in the resonator cause excess frequency fluctuations due to the effect of radiation pressure.[18] To avoid a loss of oscillator frequency stability, the power of the microwave signal incident on the resonator was stabilized. The power stabilization system[11] was based on a cryogenic amplitude detector placed near the sapphire resonator and a voltage controlled attenuator located in the room temperature part of the loop oscillator.

The amplitude detector of the power stabilization system was also used to control the bias voltage applied to phase modulator of the Pound frequency discriminator. The idea was to cancel the spurious AM-modulation of the interrogation microwave signal resulting from the dependence of phase modulator's insertion loss on applied voltage. The level of spurious AM-modulation is temperature dependent and gives rise to the temperature dependent offsets of oscillator frequency from the resonance frequency of the sapphire loaded cavity. By choosing the phase modulator with a non-monotonic dependence of the insertion loss $α$ on voltage $u$ and operating at the turning point on the dependence $α(u)$ one can eliminate the spurious AM-modulation of the interrogation signal and improve the long-term stability of the cryogenic oscillator.[19] A more detailed analysis of the servos and noise floors of the control systems will be discussed in a paper currently in preparation.

Two almost identical CSOs were constructed to measure their short-term frequency stability. The beat frequency between the two CSOs was low enough (131.181 kHz) to be directly measured with an Agilent 53132A frequency counter referenced to a 10 MHz signal from a hydrogen maser. The resulting frequency instability of a single oscillator, expressed in terms of the Allan deviation of fractional frequency fluctuations, $\sigma_y$, is shown in fig. 2. At integration times 1s < $\tau$ < 4s Allan deviation varies as $1.2 \times 10^{-15}/\sqrt{\tau}$ reaching the minimum of $5.6 \times 10^{-16}$ at 20 s. The $\sigma_y$ then increases to ~$10^{-15}$ at $10^3$ s. We estimate that the short-term frequency stability of a single CSO was limited by the intrinsic fluctuations in the electronics of the Pound frequency discriminator, namely by the noise in the cryogenic amplitude detector.

The long-term stability of the CSO was measured against a 11.2 GHz signal synthesized from a "Kvarz" hydrogen maser. A beat frequency of 386 kHz was counted with a 10 s gate time. Allan deviation of the beat note is shown in curve 4 of fig. 2, along with the frequency stability of the hydrogen maser (curve 5). The maser noise dominates these measurements at relatively short integration times for $\tau < 10^3$ s. For $\tau > 10^3$ s the CSO frequency fluctuations exceed those of a maser. In such a case, the Allan deviation of the CSO (curve 2 in fig. 2) increases with integration time as $\sigma_y \approx 2 \times 10^{-17}\sqrt{\tau}$ corresponding to the random walk of oscillator frequency. This, we believe, may be related to the ambient temperature fluctuations influencing the CSO. It is worth mentioning, that not only classical oscillators but lasers and some atomic clocks, as well, also exhibit the random walk type noise at low Fourier frequencies.[20]

It is instructive to compare the above results with those of Chang et al.[21] They quote an Allan deviation of about $5.4 \times 10^{-16}/\sqrt{\tau}$ for integration times from 1s to 4s with a minimum of $2.4 \times 10^{-16}$ within the interval 16 s < $\tau$ < 64 s. Their data were obtained with the drift removed, which was of the order of $10^{-13}$/day. Also, they only used about half an hour worth of data (inferred from the published error bars) and did not repeat their measurements. No time series data were published.[21,22] Our measurements conclusively show that short time series data segments do not reliably characterize the oscillator short term frequency instability. For example, we can select from our measured data "quiet" half hour segments and calculate an Allan deviation of $4 \times 10^{-16}$ near 20 s of averaging time.

The fractional frequency drift of the CSO reported here was measured to be $-2.2 \times 10^{-15}$/day, which is almost two orders of magnitude lower than that reported by Chang et al.[21] Such a low drift rate, we believe, was achieved due to the optimal geometry of the shielded sapphire resonator characterized by the absence of any spurious modes in the vicinity of the operational mode. This is highlighted in fig. 2 where only curve 2 has had drift removed, resulting in only a slight improvement in Allan deviation for $\tau > 2000$ s. Also the measurements were repeatable and large amounts of data were collected over many days giving consistent results (see fig. 2 caption for details). Further comparisons may be made with the LPMO oscillator[4] having a fractional frequency drift of $-6 \times 10^{-14}$/day, with the JPL oscillator characterized by a drift rate of $-10^{-14}$/day[3] and with other UWA CSOs,[1] which have negative drifts of the order of $10^{-13}$/day.

The CSO vs maser frequency comparison was conducted over the three periods of 9 (shown in fig. 3), 18 and 8 days long. From a linear fit to the 9-day data we deduce the drift rate of $-2.2 \times 10^{-15}$/day. The 8-day comparison gives us $-3.3 \times 10^{-15}$/day. The 18-day data set had two disturbances, which we believe bias the measurements. Firstly after 5 days one of stoppers on the helium dewar came out and the internal helium pressure dropped. Secondly after about 14 days some heavy objects were dropped near the dewar but the oscillator slowly recovered. If we measure the drift rate between these two events we obtain the value of $-7.7 \times 10^{-15}$/day. In all three data sets diurnal variations of the beat frequency between two CSOs were clearly seen. They are most likely due to daily temperature variations, but the further research is needed to firmly establish the causes of such variations.

We report on the measurement of the long term performance of an ultra-stable cryogenic sapphire oscillator operating at 11.200 GHz exhibiting an Allan deviation less than $10^{-15}$ for integration times between $1s < \tau < 10^3$ s and the lowest fractional frequency drift ever measured in experiments with classical microwave "clocks."

We would like to thank Dr Paul Abbott for his assistance and advice, and NICT, Japan, for providing use of the second CSO for the short-term measurements. This research was funded by the Australian Research Council.

Figure 1: Schematic on the sapphire loaded copper cavity showing the support structure and coupling probes. The important room temperature dimensions are Dc = 80 mm, Lc = 50 mm, D = 51.00 mm, and L = 30.00 mm. All dimensions are in mm.

Figure 2: Oscillator fractional frequency stability represented by the Allan deviation, $\sigma_y$, as a function of integration time, $\tau$. Curve 1 of fig. 2 is the Allan deviation calculated from 15.9 hours of data (or 57317 samples) collected with a gate time of 1 s. Curve 2 of fig. 2 is $\sigma_y$ calculated from 16.7 hours of data (or 6022 samples) collected with a gate time of 10 s. Curve 3 is the frequency counter noise floor. Curve 4 of fig. 2 is $\sigma_y$ calculated from 9 days of data (or 76333 samples, shown in fig. 3) also collected with a 10 s gate time. Curve 5 is the Kvarz hydrogen-maser stability specification, where the line represents the upper limit.

Figure 3: Beat note minus 386 kHz between the CSO and a synthesized signal from a Kvarz hydrogen-maser, measured over 9 days using a gate time of 10 s.

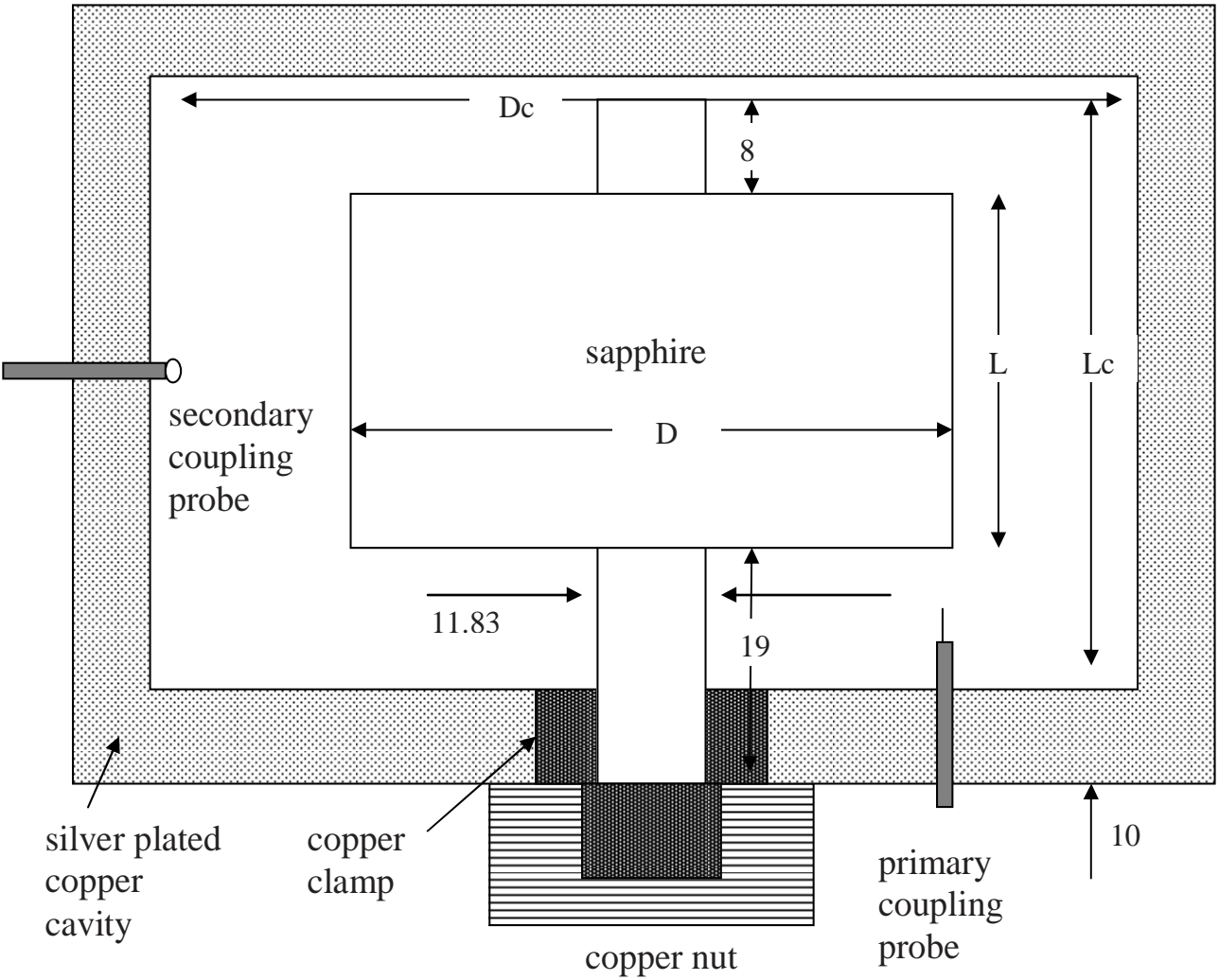

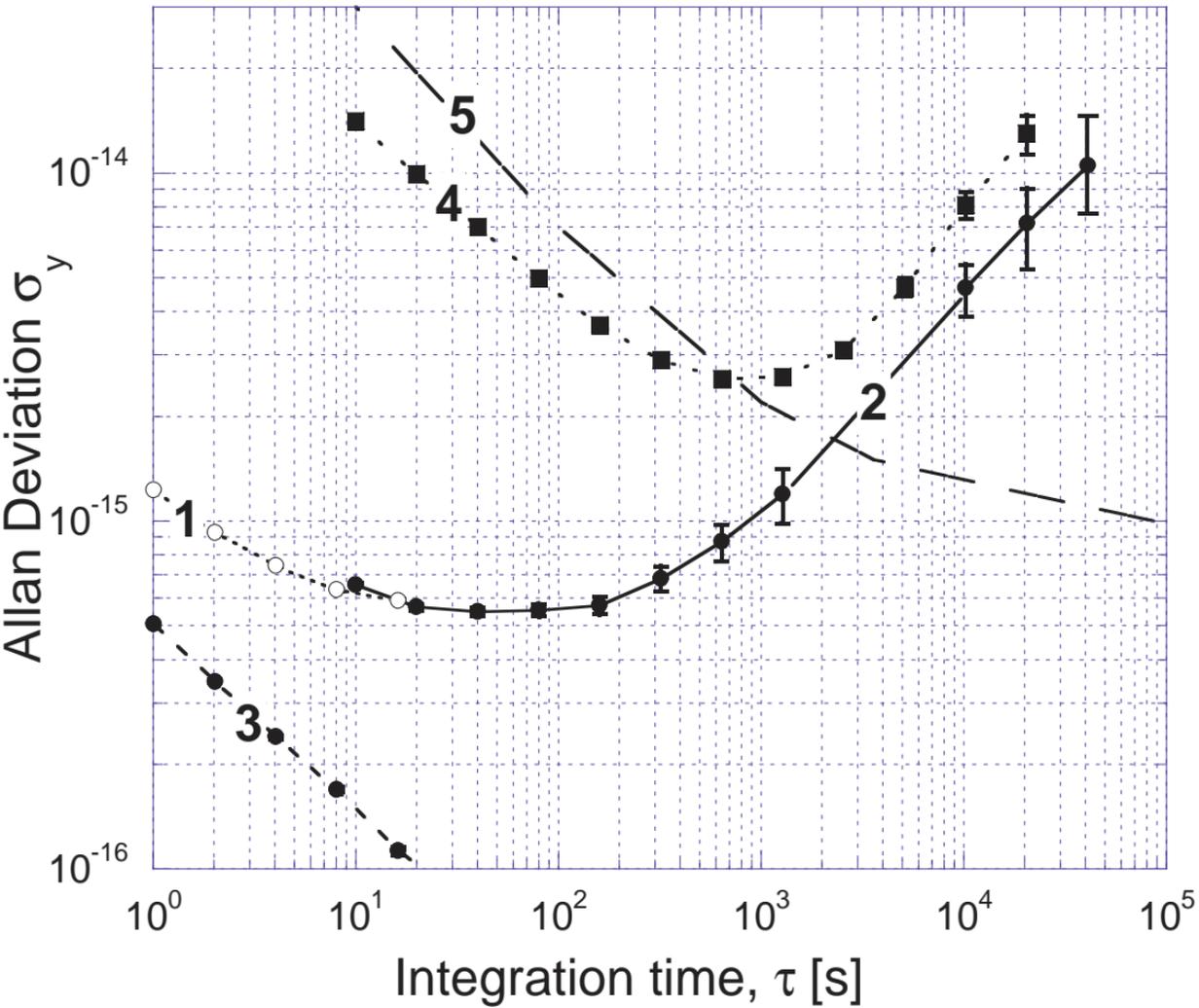

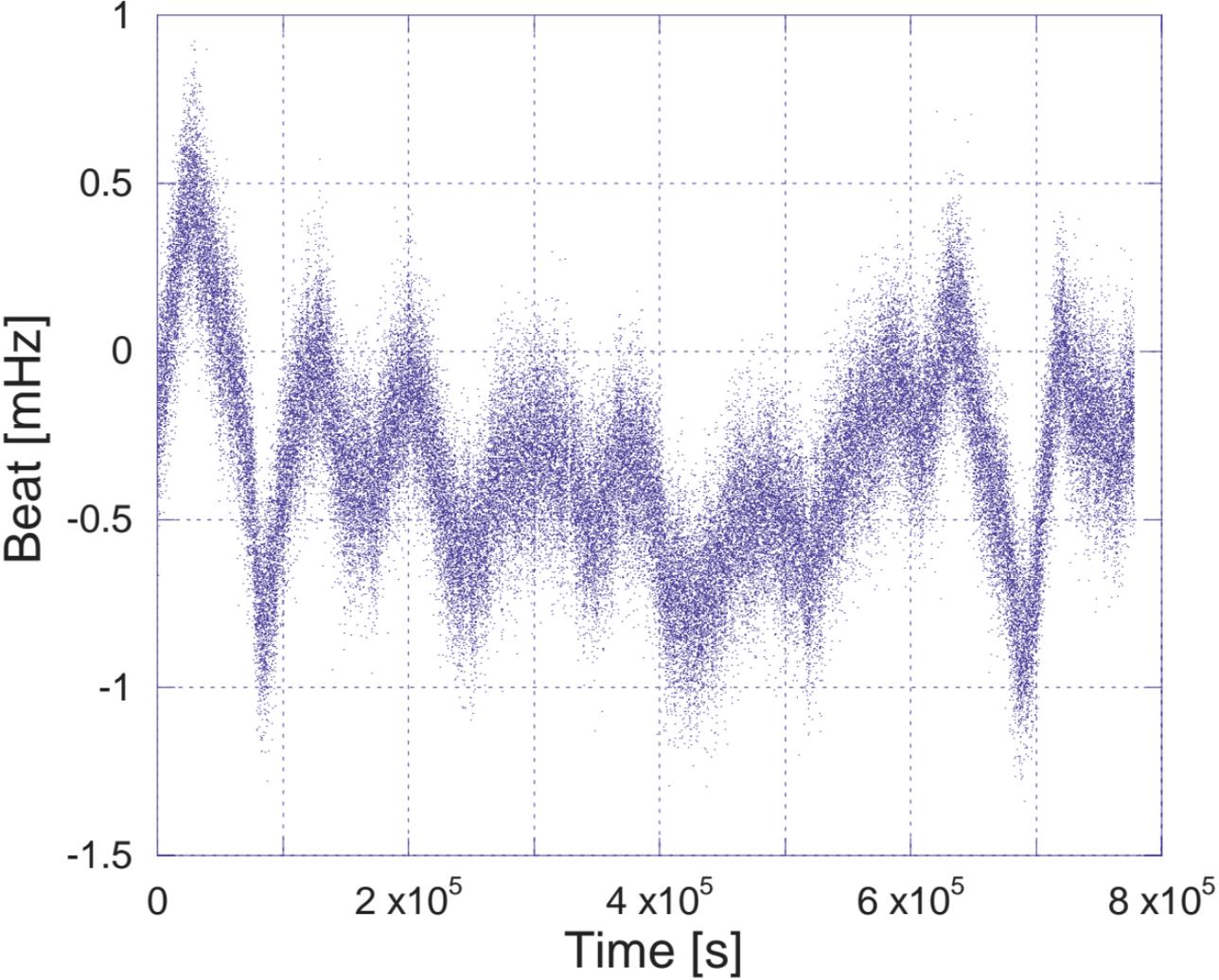